\documentclass[12pt]{article}

\usepackage{graphicx}
\usepackage{amssymb}
\usepackage{amsmath}
\usepackage{color}

\def\bea{\begin{eqnarray}}
\def\eea{\end{eqnarray}}
\def\beq{\begin{equation}}
\def\eeq{\end{equation}}

\begin{document}

\begin{center}
{\Large{\bf Kinetics of Phase Transitions in Quark Matter}} \\
\ \\
\ \\
by \\
Awaneesh Singh$^1$, Sanjay Puri$^1$ and Hiranmaya Mishra$^{1,2}$ \\
$^1$School of Physical Sciences, Jawaharlal Nehru University, \\New Delhi -- 110067, India. \\
$^2$Theory Division, Physical Research Laboratory, Navrangpura, \\Ahmedabad -- 380009, India.
\end{center}

\begin{abstract}
We study the kinetics of chiral transitions in quark matter using a phenomenological framework (Ginzburg-Landau model). We focus on the effect of inertial terms on the coarsening dynamics subsequent to a quench from the massless quark phase to the massive quark phase. The domain growth process shows a crossover from a fast inertial regime [with $L(t) \sim t (\ln t)^{1/2}$] to a diffusive Cahn-Allen regime [with $L(t)\sim t^{1/2}$].
\end{abstract}

\newpage

There has been intense interest in the {\it kinetics of phase transitions}, and the ordering process that occurs after a rapid quench in system parameters, e.g., temperature, pressure \cite{aj94,pw09}. During the transition, the system develops a spatial network of randomly-distributed domains which coarsen with time. This {\it domain growth} process has been extensively studied in magnets, alloys and fluids, liquid crystals, superconductors, the early universe, etc. \cite{aj94,pw09}. In this letter, we study an important application in quark matter, i.e., kinetics of chiral transitions.

The present study has several novel features from the perspective of both quark matter and domain growth. First, we study the properties of the coarsening morphology (e.g., correlation functions, growth laws, etc.) in chiral transitions. These universal features are independent of system and model details, and can in principle be measured in the ongoing and planned experiments on {\it quark-gluon plasma} (QGP). Second, the chiral transition provides a context to study ordering dynamics in the $\psi^6$ Ginzburg-Landau (GL) potential, which has received little attention to date. Third, chiral dynamics provides a framework to investigate the effect of inertial terms in phase ordering systems. Studies of domain growth have primarily focused on models with dissipative overdamped dynamics. We will examine all these aspects in this letter.

Heavy-ion collision experiments at high energies produce hot and dense strongly-interacting matter, and provide an opportunity to explore the phase diagram of {\it quantum chromodynamics} (QCD) in the plane of temperature ($T$) and baryon chemical potential ($\mu$). Many model studies \cite{mk98}, as well as recent lattice studies \cite{latmu}, indicate that at sufficiently large baryonic densities, there is a line of first-order transitions in the ($\mu, T$)-plane between a chirally-symmetric phase and a broken-symmetry phase. As one moves along the phase boundary towards higher $T$ and smaller $\mu$, the first-order transition becomes weaker -- ending in a {\it tricritical point} (TCP) in the limit of vanishing current quark mass or a rapid crossover for non-zero current quark mass \cite{misha}. For even smaller values of $\mu$, there is a line of second-order transitions. While the high-$T$  and small-$\mu$ region of the QCD phase diagram has been explored in recent experiments, future heavy-ion  collision experiments plan to explore the high baryon density regime, particularly the region around the TCP \cite{cpod}. 

It is important to stress here that heavy-ion experiments are essentially nonequilibrium processes. Therefore, an understanding of the equilibrium phase diagram alone is not sufficient to discuss the properties of the system. One also has to understand the kinetic processes which drive the phase transition, and the properties of the nonequilibrium structures that the system goes through to reach equilibrium. In this context, both the {\it critical dynamics} and the {\it far-from-equilibrium kinetics} of the chiral transition have attracted much recent attention.

To place our work in the proper context, let us briefly review some closely-related studies. Fraga and Krein \cite{fragaplb} undertook an important study of far-from-equilibrium kinetics in QGP. They modeled the relaxation to equilibrium via a Langevin equation, which can be derived from a microscopic field-theoretic model of order-parameter kinetics \cite{gr94,dr98}. Fraga and Krein studied the early stages of {\it spinodal decomposition} in this model, and focused upon the effect of dissipation on the spinodal instability. In recent work, Bessa et al. \cite{fraga} studied bubble {\it nucleation kinetics} in chiral transitions, and the dependence of the nucleation rate on various parameters.

Skokov and Voskresensky \cite{skokovdima} have also studied the kinetics of first-order phase transitions in nuclear systems and QGP. Starting from the equations of non-ideal non-relativistic hydrodynamics, they derived {\it time-dependent Ginzburg-Landau} (TDGL) equations for the coupled order parameters. These TDGL equations were studied numerically and analytically in the vicinity of the critical point. Skokov-Voskresensky focused upon the evolution of density fluctuations in the metastable and unstable regions of the phase diagram, and the growth kinetics of seeds. They also clarified the role of viscosity in the ordering kinetics. Finally, we mention the recent work of Randrup \cite{jr09}, who studied the fluid dynamics of relativistic nuclear collisions. The corresponding evolution equations reflect the conservation of baryon charge, momentum and energy. Randrup studied the amplification of spinodal fluctuations and the evolution of the {\it correlation function}. Randrup's work mostly focused upon the evolution in the linearized regime, where there is an exponential growth of initial fluctuations.

This letter and a recent companion paper \cite{spm10} are complementary to Refs.~\cite{fragaplb,fraga,skokovdima,jr09}. Our study investigates the late stages of phase-separation kinetics in quark matter and the scaling properties of emergent morphologies. The system is described by nonlinear evolution equations in this regime: the exponential growth of initial fluctuations is saturated by the nonlinearity. We consider an initially disordered system which is rendered thermodynamically unstable by a rapid quench to the broken-symmetry phase. We study domain growth and highlight quantitative features of the coarsening morphology. We also study the evolution kinetics of single droplets, and the dependence of the front velocity on system parameters.

To model chiral symmetry breaking in QCD, we use the two-flavor Nambu-Jona-Lasinio (NJL) model \cite{klevansky,glfree} with the Hamiltonian:
\begin{eqnarray}
H = \sum_{i,a}\psi^{ia \dagger}\left(-i\vec{\alpha}\cdot\vec{\nabla} + \gamma^0 m_i \right)\psi^{ia}-G\left[(\bar\psi\psi)^2-(\bar\psi\gamma^5 \tau \psi)^2\right].
\label{ham}
\end{eqnarray}
Here, $m_i$ is the current quark mass -- we take this to be the same ($m_i=m$) for both $u$ and $d$ quarks. The parameter $G$ denotes the quark-quark interaction strength. Further, $\tau$ is the Pauli matrix acting in flavor space. The quark operator $\psi$ has two indices $i$ and $a$, denoting the flavor and color indices, respectively.
To describe the ground state, we take an ansatz with arbitrary 
number of quark-antiquark pairs as \cite{hmnj}:
\begin{equation} 
|vac\rangle= \exp\!\!\left[\!\int\!\! d\vec{k}~ q_I^{0}(\vec{k})^\dagger(\vec{\sigma}\cdot\vec{k})h(\vec{k})\tilde q_I^{0} (-\vec{k})-\mathrm{h.c.}\right]\!|0\rangle.
\label{uq}
\end{equation}
\textcolor{red}{Here, $q_I^{0\dagger}$, $\tilde q_I^0$ are two-component quark and antiquark creation operators, and $|0\rangle$ is the perturbative chiral vacuum (i.e., $q_I^0|0\rangle=0=\tilde q_I^{0\dagger}|0\rangle$). Further, $h(\vec{k})$ is a variational function related to the condensate as $\langle\bar{\psi}\psi\rangle = -12 \int \!d\vec{k}\,\sin[2h(\vec{k})]/(2 \pi^3)$. The prefactor $12 = 2 \times 3 \times 2$ arises from flavor, color and spin degrees of freedom. A nontrivial $h(\vec k)$ clearly breaks chiral symmetry.}

This condensate function $h(\vec k)$ can be determined by minimizing the energy at $T=0$, or the thermodynamic potential at non-zero $T$ and $\mu$. In Ref.~\cite{spm10}, we have obtained the thermodynamic potential as a function of $T$ and $\mu$:
\begin{align}
\tilde\Omega(M,\beta,\mu) =& \;-\dfrac{12}{(2\pi)^3\beta}\displaystyle\int \! d\vec{k}\;\Big\{ \ln\left[1+e^{-\beta\left( \sqrt{k^2+M^2} - \mu \right)}\right] \notag\\
& \qquad\qquad\qquad\quad + \ln\left[1+e^{-\beta\left( \sqrt{k^2+M^2} + \mu \right)}\right] \Big\}   \notag\\
&\; -\dfrac{12}{(2\pi)^3}\displaystyle\int \! d\vec{k} \; \left(\sqrt{k^2+M^2}-k\right)+ \dfrac{M^2}{4G} ,
\label{tomega}
\end{align}
where $\beta = (k_B T)^{-1}$. Here, we have taken vanishing current quark mass ($m=0$), and introduce $M=-2g\rho_s$, with $\rho_s=\langle\bar\psi\psi\rangle$ being the scalar density and $g=G[1+1/(4N_c)]$.

The potential in Eq.~(\ref{tomega}) may be expanded as a Landau potential in the order parameter $M$: 
\begin{equation}
\tilde\Omega\left(M \right)= \tilde\Omega\left(0 \right) + \frac{a}{2}M^2 + \frac{b}{4}M^4 + \frac{d}{6}M^6 + O(M^8) \equiv f(M) ,
\label{p6}
\end{equation}
correct up to logarithmic factors \cite{glfree}. In the following, we consider the expansion of $\tilde\Omega\left(M \right)$ up to the $M^6$-term. This will prove adequate to \textcolor{red}{capture the proposed phase structure of QCD, as we see shortly}. The first two coefficients in Eq.~(\ref{p6}) can be obtained by comparison with Eq.~(\ref{tomega}) as
\begin{align}
\tilde\Omega(0) =&\;-\dfrac{6}{\pi^2\beta}\displaystyle\int_0^\Lambda \!\!\! dk\,\, k^2 \left\lbrace  
\ln\left[1+e^{-\beta(k-\mu)}\right] + \ln\left[1+e^{-\beta(k+\mu)}\right]\right\rbrace, \nonumber \\
a =& \; \dfrac{1}{2G} - \dfrac{3\Lambda^2}{\pi^2} + \dfrac{6}{\pi^2}\displaystyle\int_0^\Lambda \!\!\! dk\,\,k\left[ \dfrac{1}{1+e^{\beta(k-\mu)}} + \dfrac{1}{1+e^{\beta(k+\mu)}}\right]. 
\label{coff}
\end{align}
We treat the higher coefficients as phenomenological parameters, which are obtained by fitting $\tilde\Omega\left(M \right)$ in Eq.~(\ref{p6}) to the integral expression for $\tilde\Omega (M)$ in Eq.~(\ref{tomega}) \cite{spm10}. There are two free parameters in the microscopic theory ($\mu$ and $T$), so we consider the $M^6$-Landau potential with parameters $b$ and $d$. For stability, we require $d>0$.

\begin{figure}[!htb]
\centering
\includegraphics[width=0.7\textwidth]{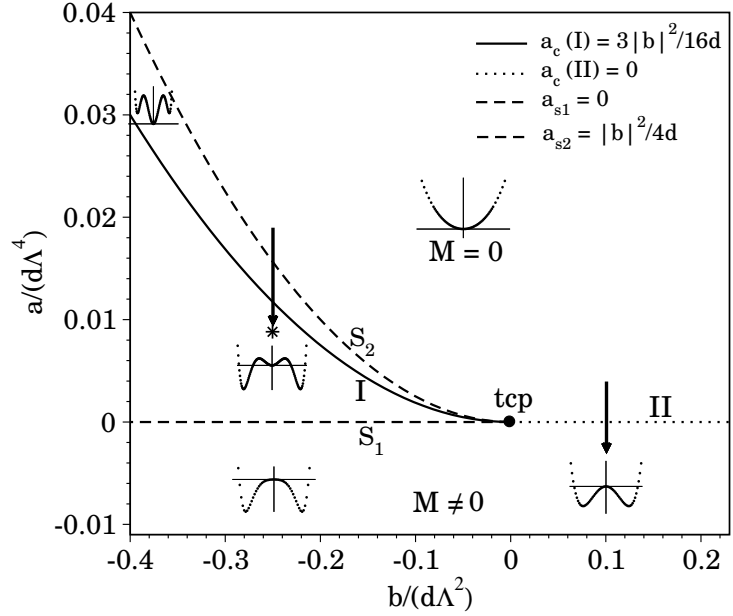}
\caption{Phase diagram for the Landau free energy in Eq.~(\ref{p6}) in the [$b/(d\Lambda^2), a/(d\Lambda^4)$]-plane. A line of first-order transitions (I) meets a line of second-order transitions (II) at the tricritical point (tcp), which is located at $a=b=0$. The equations for I and II are specified in the figure. The dashed lines denote the spinodals $S_1$ and $S_2$, whose equations are also provided. The typical forms of the Landau potential in various regions are shown in the figure. The asterisk denotes the point where we quench the system for $b<0$ (first-order quench). The second-order quench studied here corresponds to $b/(d\Lambda^2) = 1.269$, $a/(d\Lambda^4) = -0.225$, and is not shown in the figure for clarity.}
\label{fig1}
\end{figure}

The extrema of the potential in Eq.~(\ref{p6}) are determined by the gap equation: $f'(M)=aM+bM^3+dM^5=0$. The corresponding solutions are $ M=0$, and $M_{\pm}^2=(-b\pm \sqrt{b^2 -4ad})/(2d)$. The phase diagram for the Landau potential is shown in Fig.~\ref{fig1}. (A) For $b>0$, the transition is second-order, analogous to an $M^4$-potential -- the stationary points are $M=0$ (for $a>0$) or $M=0$, $\pm M_+$ (for $a<0$). For $a<0$, the preferred equilibrium state is the one with massive quarks. (B) For $b<0$, the solutions of the gap equation are as follows: (i) $M=0$ for $a>|b|^2/(4d)$, (ii) $M=0$, $\pm M_+$, $\pm M_-$ for $|b|^2/(4d)>a>0$, and (iii) $M=0$, $\pm M_+$ for $a<0$. A first-order transition takes place at $a_c=3|b|^2/(16d)$ with the order parameter jumping discontinuously from $M=0$  to $M=\pm M_+=\pm (3|b|/4d)^{1/2}$. The tricritical point is located at $b_\text{tcp}=0$, $a_\text{tcp}=0$.

Next, we study dynamical problems in the context of the above free energy. Consider the environment of a heavy-ion collision. If the evolution is slow compared to the typical equilibration time, the order parameter field will be in local equilibrium. We consider a system which is rendered thermodynamically unstable by a rapid quench from the massless phase to the massive phase in Fig.~\ref{fig1}. The unstable massless state evolves via the emergence and growth of domains rich in the preferred massive phase \cite{aj94,pw09}. The coarsening system is inhomogeneous, and we account for this by including a surface tension term in the Landau free energy:
\begin{equation}
\Omega[M]=\int d\vec{r} \left[\frac{a}{2} M^2+\frac{b}{4} M^4+\frac{d}{6} M^6 +
\frac{K}{2}\left(\vec{\nabla} M\right)^2\right] .
\label{omgl}
\end{equation}

It is customary to model the kinetics by the TDGL equation, which models the overdamped (relaxational) dynamics of an order-parameter field to the minimum of the potential in Eq.~(\ref{omgl}). The inertial term with a second-order time-derivative is usually neglected in comparison to the damping term which is first-order in the time-derivative. We have studied the ordering dynamics of such a TDGL model in Ref.~\cite{spm10}.

However, a microscopic derivation of the kinetic equation in a relativistic field theory using, e.g., the {\it closed-time-path Green's function}  (CTPGF) formalism results in a second-order stochastic equation. Such a derivation has been done for scalar field theories \cite{gr94,dr98,bd99}. A second-order TDGL equation has also been derived for the NJL model by Fu et al. \cite{fh11} using the CTPGF method. More recently, a Langevin equation with an inertial term has been derived for the chiral order parameter field in a sigma model by Nahrgang et al. using an influence-functional method \cite{nl11}. This model has been used to discuss the relaxational dynamics of the order parameter near the critical point \cite{nahrgang2,nahrgang3}. Given this background, it is relevant to investigate the effect of an inertial term on the ordering kinetics of the chiral transition. More generally, it is important to study the effect of an inertial term in domain growth problems. In spite of the intense interest in this area, this question has received almost no attention \cite{aj94,pw09}. In this letter, we will address this issue in the context of chiral transitions.

Therefore, we consider a system whose evolution is described by the TDGL equation with an inertial term: 
\begin{equation}
\frac{\partial^2}{\partial t^2} M (\vec{r}, t)+ \bar{\gamma} \frac{\partial M}{\partial t} = -\frac{\delta \Omega\left[M \right]}{\delta M(\vec r,t)} + \theta\left(\vec{r},t\right) ,
\label{ke}
\end{equation}
where $\bar{\gamma}$ is the dissipation coefficient. Here, $\theta(\vec{r},t)$ is the noise term with zero average, satisfying the fluctuation-dissipation relation ($k_B=1$): $\langle \theta(\vec{r'},t')\theta(\vec{r''},t'') \rangle = 2 \bar{\gamma} T\delta(\vec{r'}-\vec{r''})\delta\left(t'-t''\right)$. We use the natural scales of order parameter, space and time to introduce dimensionless variables: 
\begin{eqnarray}
M &=& M_0 M', \quad M_0=\sqrt{|a|/|b|} , \nonumber \\
\vec{r} &=& \xi \vec{r'}, \quad \xi=\sqrt{K/|a|} , \nonumber \\
t &=& \tau t', \quad \tau= 1/\sqrt{|a|} , \nonumber \\
\theta &=& (|a|^{3/2}/|b|^{1/2})~\theta' .
\end{eqnarray}
Dropping the primes, we obtain the dimensionless TDGL equation:
\begin{eqnarray}
\frac{\partial^2 M}{\partial t^2} + \gamma\frac{\partial M}{\partial t} = -\mathrm{sgn} (a) M -\mathrm{sgn} (b) M^3 - \lambda M^5 + \nabla^2 M +\theta\left(\vec{r},t\right),
\label{ke2}
\end{eqnarray}
where $\gamma = \bar{\gamma}/\sqrt{|a|}$, $\mathrm{sgn} (x) = x/|x|$, and $\lambda=|a|d/|b|^2 >0$. The dimensionless noise satisfies
\bea
\left\langle \theta(\vec{r'},t')\theta(\vec{r''},t'') \right\rangle &=& 2\epsilon\delta(\vec{r'}-\vec{r''})
\delta\left(t'-t''\right) , \nonumber \\
\epsilon &=& \frac{\gamma T |b|}{|a|^{(5-d)/2} K^{d/2}} ,
\label{fdr}
\eea
where $d$ is the dimensionality.

Our results in this letter are presented in dimensionless units of 
space and time. To obtain these in physical units, one has to multiply
 by the appropriate dimensional quantities $\xi$ and $\tau$. For this,
 we need to estimate the strength of the interfacial energy $K$. 
The surface tension can be calculated as $\sigma = \sqrt{K}(|a|^{3/2}/|b|)
\int dz(dM_s/dz)^2$, where $M_s(z)$ is the static kink solution of
 Eq.~(\ref{ke2}) with $\theta = 0$. For quark matter, $\sigma$ is poorly 
known and varies from 10-100 MeV/$\text{fm}^2$ at small temperatures
 \cite{hc93}. \textcolor{red}{On the other hand, recent estimates using effective models \cite{signjl} like the NJL model and the Polyakov loop-quark-meson model suggest a lower value for surface tension: $\sigma\simeq 5-20$MeV. We take $\sigma \simeq 10$ MeV/$\text{fm}^2$.} For $T=10$ MeV and $\mu = 321.75$ MeV, we then estimate $\xi = \sqrt{K/|a|} \simeq 0.56$ fm and $\tau =  1/\sqrt{|a|} \simeq 5.1$ fm \cite{spm10}.

We study the phase transition kinetics for two different quench possibilities. First, we consider deep quenches through II ($b>0$) from $a>0$ (with $M=0$) to $a<0$, where the free energy has a double-well structure. \textcolor{red}{Notice that we quench far below the line of second-order transitions.} The chirally-symmetric phase is now unstable, and evolves to the stable massive phase via spinodal decomposition. In our simulations of this case, we have used Eq.~(\ref{ke2}) with $a<0, b>0, \lambda=0.14$, corresponding to $(\mu, T) =$ (231.6 MeV, 85 MeV) \cite{spm10}:
\begin{align}
\frac{\partial^2 M}{\partial t^2} + \gamma\frac{\partial M}{\partial t} = M - M^3 - \lambda M^5 + \nabla^2 M +\theta\left(\vec{r},t\right).
\label{ke3}
\end{align}
We solve Eq.~(\ref{ke3}) numerically using a simple Euler-discretization scheme with initial velocity $\partial M/\partial t |_{t=0} = 0$. The initial state of the system is prepared as  $M(\vec{r},0) = 0 \pm \delta M(\vec{r},0)$, where $\delta M$ is uniformly distributed in [$-0.25, +0.25$].

Our numerical simulations are performed on a $d=3$ lattice of size $N^3$ ($N=256$), with periodic boundary conditions in all directions. \textcolor{red}{The discretization mesh sizes are $\Delta x = 1.0$ and $\Delta t = 0.1$, obtained from the linear stability analysis of Eq.~(\ref{ke3}) \cite{yp87,red88}. We require that the Euler scheme must respect the stability properties of the homogeneous solutions of Eq.~(\ref{ke3}), and not contain unphysical divergences.} The thermal noise $\theta(\vec{r},t)$ is mimicked by uniformly-distributed random numbers between $[-A_n, A_n]$. \textcolor{red}{In studies of phase-transition kinetics, it is known that statistical results are unchanged whether we use Gaussian noise or uniformly-distributed noise \cite{yp87,po88}.} The appropriate noise amplitude in our simulation is \cite{sp02} $A_n = \sqrt{3\epsilon/(\Delta x^d \Delta t)}$. The results reported here correspond to $\epsilon=0.008$, i.e., $A_n=0.5$. All statistical quantities are obtained as averages over 10 independent runs.

\begin{figure}[!htb]
\centering
\includegraphics[width=0.9\textwidth]{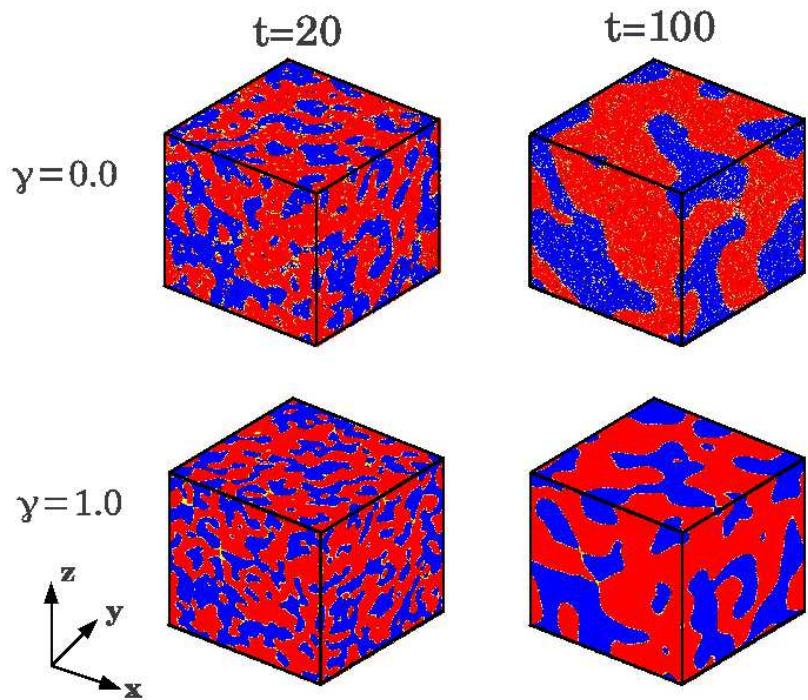}
\caption{Domain growth for $\gamma = 0.0, 1.0$ after a quench through the second-order line (II) in Fig.~\ref{fig1}. The snapshots show regions with $M \simeq +M_+$ (marked blue), $M \simeq 0$ (marked yellow), and $M \simeq -M_+$ (marked red) at $t=20, 100$. The simulation details are provided in the text.}
\label{fig2}
\end{figure}

In Fig.~\ref{fig2}, we show the ordering dynamics of Eq.~(\ref{ke3}) from a disordered initial state. To study the effect of inertia, we chose $\gamma=0.0$ (upper frames) and 1.0 (lower frames). The system rapidly evolves into domains of the massive phase with $M\simeq M_+$ (marked red) and $M\simeq -M_+$ (marked blue). The snapshots show the evolution at $t=20, 100$. For $\gamma=0$, the dissipative term is absent, and we observed a rapid growth of domains (see the pattern at $t=20$). After the initial rapid growth, domain walls get fuzzier, and domains become less distinctive due to the oscillatory behavior of the system. We have also studied the time-dependence of the order-parameter value at a few spatial points in the $\gamma = 0$ case. We observe the occurrence of flips from $\pm M_+ \rightarrow \mp M_+$ on extended time-scales. In spite of these, the domain morphology continues to coarsen as these oscillations are cooperative. For $\gamma = 1$, the dissipative term is dominant and the ordering dynamics is analogous to that for the overdamped case \cite{spm10}.

\begin{figure}[!htb]
\centering
\includegraphics[width=0.7\textwidth]{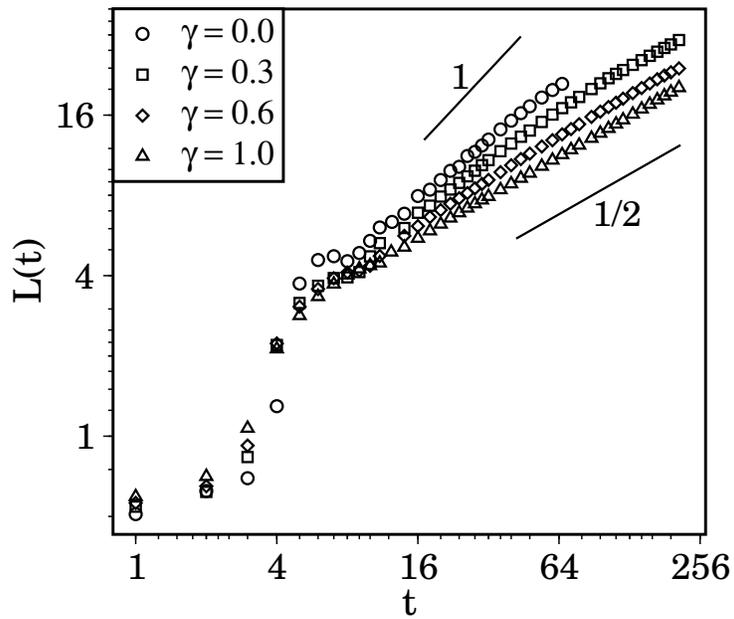}
\caption{Time-dependence of domain size, $L(t)$ vs. $t$, for the evolution depicted in Fig.~\ref{fig2}. There is a crossover at $t_c \sim \gamma^{-1}$ from an early-time inertial growth [$L(t) \sim t (\ln t)^{1/2}$] to a late-time Cahn-Allen (CA) growth [$L(t) \sim t^{1/2}$].} 
\label{fig3}
\end{figure}

The system is characterized by a single length scale $L(t)$ as the pattern morphology does not change in time apart from a scale factor. We have confirmed numerically (not shown here) that the correlation functions at different times obey {\it dynamical scaling} \cite{aj94,pw09} for different $\gamma$-values. We will present results for the scaling of the correlation function in a separate publication. Here, we focus on the time-dependence of the length scale. \textcolor{red}{The length scale $L(t)$ is defined as the distance over which the correlation function decays to half its maximum value [$C(r,t)=1$ at $r =0$].} In Fig.~\ref{fig3}, we plot $L(t)$ vs. $t$ on a log-log scale for several values of $\gamma$. As usual, $L(t)$ shows a power-law behavior [$L(t) \sim t^\phi$], but there is a distinct crossover in the exponent $\phi$ as $\gamma$ is varied.

To understand this, we consider the deterministic version ($\theta = 0$) of Eq.~(\ref{ke3}), which we rewrite as
\beq
\frac{\partial^2 M}{\partial t^2} + \gamma\frac{\partial M}{\partial t} = -f'(M) + \nabla^2 M ,
\label{ke3d}
\eeq
where 
\beq
f(M) = -\frac{M^2}{2} + \frac{M^4}{4} + \lambda \frac{M^6}{6} .
\label{poten}
\eeq
The 1-dimensional static (kink) solution $M_s(z)$ of Eq.~(\ref{ke3d}) is the same in the inertial and overdamped cases and obeys
\beq
-f'(M_s) + \frac{d^2 M_s}{dz^2} = 0 .
\label{ms}
\eeq
Equation~(\ref{ms}) yields a tanh-profile between $M=-1$ and $M=+1$ for the usual $M^4$-potential: $f(M) = -M^2/2 + M^4/4$. The corresponding kink profile for the potential in Eq.~(\ref{poten}) connects the two vacuum states: $+M_+$ and $-M_+$, where $M_+^2 = (-1+ \sqrt{1+4 \lambda})/(2\lambda)$.

For Eqs.~(\ref{ke3d})-(\ref{poten}), we consider a droplet of $M=+M_+$ shrinking in a background with $M=-M_+$. If the radius of the droplet is $R(t)$, then
\beq
M(\vec{r},t) \simeq h[r-R(t)] \equiv h(\eta) ,
\label{sig}
\eeq
where \textcolor{red}{$r = |\vec{r}|$}, and $h(\eta)$ is a sigmoidal profile whose derivative is sharply peaked at $r=R(t)$. Replacing Eq.~(\ref{sig}) in Eq.~(\ref{ke3d}), we obtain
\beq
0 = h'' \left[ 1 - \left(\frac{dR}{dt}\right)^2 \right] + h' \left( \frac{d-1}{r} + \gamma \frac{d R}{dt}
+ \frac{d^2 R}{dt^2} \right) - f'(h) .
\label{hpp2}
\eeq
We multiply Eq.~(\ref{hpp2}) by $h'$ and integrate through the interface. The first term on the RHS drops out because $h' = 0$ as $\eta \rightarrow \pm \infty$, and the third term drops out because $f(M_+) = f(-M_+)$. This yields the kinetic equation for droplet shrinkage:
\beq
\frac{d^2 R}{dt^2} + \gamma \frac{d R}{dt} = - \frac{d-1}{R} .
\eeq

The analogous growth equation for the characteristic length scale $L(t)$ is \cite{aj94,pw09}
\beq
\frac{d^2 L}{dt^2} + \gamma \frac{d L}{dt} = \frac{\sigma}{L} ,
\eeq
where $\sigma/L$ is identified as the curvature for a domain of size $L$. At short times ($t \ll t_c$), the growth law is fixed by the inertial term as \cite{bo84}
\beq
L(t) \sim \sqrt{\sigma} t \left[ \ln (\sqrt{\sigma} t) \right]^{1/2} .
\eeq
The long-time ($t \gg t_c$) kinetics is determined by the dissipative term as
\beq
L(t) \sim \left( \frac{\sigma t}{\gamma} \right)^{1/2} ,
\eeq
which is the usual Cahn-Allen (CA) growth law \cite{pw09}. The crossover time scales as $t_c \sim \gamma^{-1}$. In Fig.~\ref{fig3}, we have plotted straight lines corresponding to $L(t) \sim t$ and $L(t) \sim t^{1/2}$, the two limiting behaviors of the growth law.

Second, we consider shallow quenches through I to the point marked by a asterisk in Fig.~\ref{fig1}. This case is studied using Eq.~(\ref{ke2}) with $a>0, b<0, \lambda=0.14$, which is equivalent to $(\mu, T)$ = (321.75 MeV, 10 MeV) \cite{spm10}. The corresponding kinetic equation is
\beq
\frac{\partial^2 M}{\partial t^2} + \gamma\frac{\partial M}{\partial t} = -M + M^3 - \lambda M^5 + \nabla^2 M +\theta\left(\vec{r},t\right).
\label{ke4}
\eeq
The initial state with massless quarks ($M=0$) is now a metastable state of the potential, and phase separation proceeds via nucleation and growth of droplets of $M = \pm M_+$. Therefore, $\theta(\vec{r},t)$ must be sufficiently large to enable the system to escape from the metastable state on a reasonable time-scale: a suitable value for $\lambda = 0.14$ is $\epsilon = 0.6$. However, the asymptotic behavior of domain growth in both the unstable and metastable cases is insensitive to the noise term \cite{po88}.

\begin{figure}[!htb]
\centering
\includegraphics[width=0.85\textwidth]{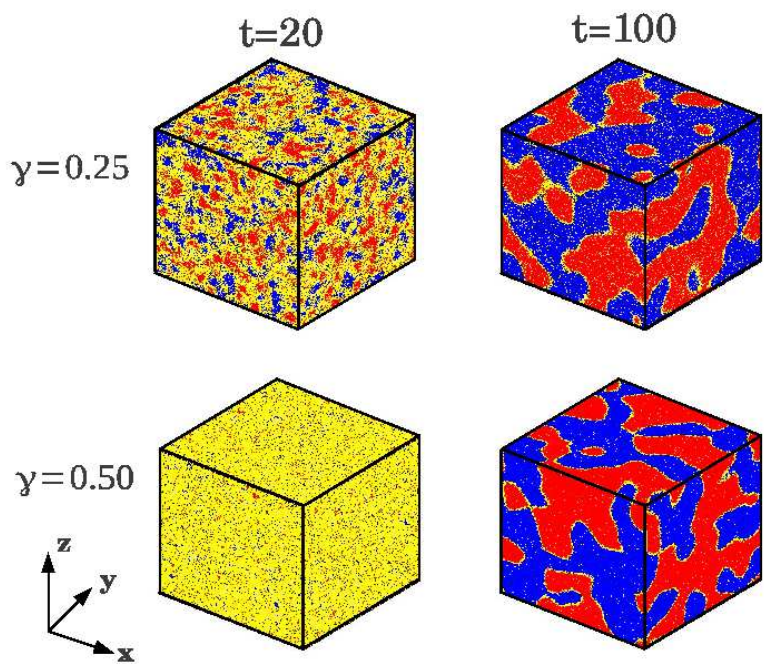}
\caption{Analogous to Fig.~\ref{fig2} but for a shallow quench through the first-order line (I) in Fig.~\ref{fig1}. Notice that the metastable patches ($M \simeq 0$, marked yellow) at $t=20$ are absent at later times.}
\label{fig4}
\end{figure}

In Fig.~\ref{fig4}, we show the ordering kinetics of Eq.~(\ref{ke4}) for $\gamma = 0.25, 0.5$. Typically, the evolution of the system begins with the nucleation of droplets in the early stages: droplets larger than a critical size $R_c$ (supercritical) grow, whereas those with $R<R_c$ (subcritical) shrink. In the present simulation, the critical radius of the bubble $R_c \simeq 8$ dimensionless units. \textcolor{red}{If we convert this into physical units, $R_c \simeq 4.5$ fm.}

The droplets grow very rapidly and fuse to form bi-continuous domain structures, a characteristic of late-stage domain growth. The effect of dissipation on nucleation and growth can be understood by comparing the evolution patterns at different $\gamma$-values. The system takes more time to nucleate for extreme $\gamma$-values (i.e., $\gamma \rightarrow 0$ and $\gamma \rightarrow \infty$). To understand this behavior, we follow Hanggi's discussion \cite{ph86} of {\it Kramer's escape problem} for a barrier. Hanggi studies the crossover time from $M=0$ (the metastable state) to $M=M_+$ (the stable state) in the homogeneous version of Eq.~(\ref{ke4}). This crossover time is proportional to the nucleation time $t_n$ in our domain growth problem. We designate $\omega_b$ as the natural vibration frequency about the barrier location ($M_-$). For moderate to large dissipation ($\gamma \gg \omega_b$), the nucleation time
\beq
t_n \sim \left(\sqrt{\frac{\gamma^2}{4} + \omega_b^2} - \frac{\gamma}{2} \right)^{-1} ,
\eeq
so that $t_n \sim \gamma$ as $\gamma \rightarrow \infty$. For small dissipation ($\gamma \ll \omega_b$), we have
$t_n \sim \gamma^{-1}$, so that $t_n \rightarrow \infty$ as $\gamma \rightarrow 0$. Subsequent to nucleation, the intermediate and asymptotic growth regimes are similar to those described for spinodal decomposition, i.e., a crossover from $L(t) \sim t (\ln t)^{1/2}$ to $L(t) \sim t^{1/2}$.

In summary, we have studied the kinetics of chiral phase transitions in QCD subsequent to sudden changes in system parameters. To understand the kinetics, we must first obtain the phase diagram. In terms of the quark degrees of freedom, the phase diagram is obtained in the $(\mu, T)$-plane using the Nambu-Jona-Lasinio (NJL) model \cite{spm10}. An equivalent coarse-grained description is obtained from an $M^6$-Landau free energy.

The chiral kinetics is modeled via the nonlinear TDGL equation, and we consider both the overdamped \cite{spm10} and inertial cases. In this letter, we focus on the effect of an inertial term on ordering dynamics. We study quenches through the first-order (I) or second-order (II) transition lines in Fig.~\ref{fig1}. For quenches through II and deep quenches through I, the massless phase is spontaneously unstable and evolves to the massive phase via spinodal decomposition. For shallow quenches through I, the massless phase is metastable and the chiral transition proceeds via the nucleation and growth of droplets of the massive phase. The merger of these droplets results in late-stage domain growth similar to that for the unstable case. In all cases, the asymptotic growth process exhibits dynamical scaling, and the growth law is $L(t)\sim t^{1/2}$. The inertial term gives a pre-asymptotic regime of faster growth with $L(t)\sim t (\ln t)^{1/2}$, and the crossover time to $t^{1/2}$-growth scales as $t_c \sim \gamma^{-1}$, where $\gamma$ is the dissipation constant. Given the dynamical universality of these processes, our results are of much wider applicability than the underlying NJL Hamiltonian. Such quenches should be realizable in high-energy heavy-ion collisions. We hope that our theoretical results will catalyse appropriate experimental interest.

\newpage

\end{document}